\documentclass[twocolumn,showpacs,,usortedaddress, superscriptaddress, preprintnumbers,amsmath,amssymb]{revtex4}

\usepackage{graphicx}
\usepackage{dcolumn}
\usepackage{bm}

\begin{document}

\title{High-voltage nanosecond pulses in a low-pressure radiofrequency discharge}

\author{M.Y. Pustylnik}
\email{pustylnik@mpe.mpg.de}
\affiliation{Max-Planck-Institut f\"{u}r Extraterrestrische Physik, Giessenbachstrasse 1, 85741 Garching, Germany}
\author{L. Hou}
\affiliation{Max-Planck-Institut f\"{u}r Extraterrestrische Physik, Giessenbachstrasse 1, 85741 Garching, Germany}
\author{A.V. Ivlev}
\affiliation{Max-Planck-Institut f\"{u}r Extraterrestrische Physik, Giessenbachstrasse 1, 85741 Garching, Germany}
\author{L.M. Vasilyak}
\affiliation{Joint Institute for High Temperatures, Russian Academy of Sciences, Izhorskaya 13/19, 125412 Moscow, Russia}
\author{L. Cou\"{e}del}
\affiliation{Laboratoire de Physique des Interactions Ioniques et Mol\'{e}culaires, Centre National de la Recherche Scientifique, Aix-Marseille-Universit\'{e}, 13397 Marseille Cedex 20, France}
\author{H.M. Thomas}
\affiliation{Max-Planck-Institut f\"{u}r Extraterrestrische Physik, Giessenbachstrasse 1, 85741 Garching, Germany}
\author{G.E. Morfill}
\affiliation{Max-Planck-Institut f\"{u}r Extraterrestrische Physik, Giessenbachstrasse 1, 85741 Garching, Germany}
\author{V.E. Fortov}
\affiliation{Joint Institute for High Temperatures, Russian Academy of Sciences, Izhorskaya 13/19, 125412 Moscow, Russia}

\date{}

\begin{abstract}
An influence of a high-voltage ($3-17$~kV) $20$~ns pulse on a weakly-ionized low-pressure ($0.1-10$~Pa) capacitively-coupled radiofrequency (RF) argon plasma is studied experimentally. The plasma evolution after pulse exhibits two characteristic regimes: a bright flash, occurring within $100$~ns after the pulse (when the discharge emission increases by 2-3 orders of magnitude over the steady-state level), and a dark phase, lasting a few hundreds $\mu$s (when the intensity of the discharge emission drops significantly below the steady-state level). The electron density increases during the flash and remains very large at the dark phase. 1D3V particle-in-cell simulations qualitatively reproduce both regimes and allow for detailed analysis of the underlying mechanisms. It is found that the high-voltage nanosecond pulse is capable of removing a significant fraction of plasma electrons out of the discharge gap, and that the flash is the result of the excitation of gas atoms, triggered by residual electrons accelerated in the electric field of immobile bulk ions. The secondary emission from the electrodes due to vacuum UV radiation plays an important role at this stage. High-density plasma generated
 during the flash provides efficient screening of the RF field (which sustains the steady-state plasma). This leads to the electron cooling and, hence, onset of the dark phase.
\end{abstract}

 \maketitle

\section{Introduction}
The high-voltage nanosecond pulses are widely used in modern low-temperature plasma physics research and technology. At high pressures (of the order of $10^4-10^5$~Pa) they, e.g., provide ionization for fast plasma switches and pumping for powerful pulsed gas lasers \cite{KorMes}, or generate plasmas for biomedical applications \cite{BioMed}. In the pressure range of $10^2-10^4$~Pa such pulses are able to launch the so-called fast ionization waves propagating at a speed comparable to the speed of light \cite{FIW}, which makes it possible to use them for the fast ignition of chemically reactive gas mixtures \cite{Ignition}. High-voltage pulses can also stabilize discharges in powerful CO$_2$ lasers \cite{Laser}.

 At the same time, there is a growing interest in applying the high-voltage nanosecond pulses under low-pressure conditions, i.e. in the range of $10^{-1}-10^2$~Pa. Amirov et. al. \cite{Amirov1, Amirov2}, who were the first to combine a classical dc glow discharge in a glass tube with a short high-voltage pulse, reported on the so-called ``glow pause'' (or ``dark phase'', as it was termed later in Refs. \onlinecite{DP1, DP2})~\textendash~a period of time after the pulse when the discharge becomes practically dark. Later, several experiments were reported, in which high-voltage nanosecond pulses were used to manipulate the dust particles levitating in weak low-pressure discharges [\onlinecite{NanoVas1}-\onlinecite{NanoPust2}]. Another idea was to combine capacitively-coupled radiofrequency (RF) and high-voltage nanopulse discharges, in order to enhance the production of H$^-$ ions in low-pressure hydrogen plasmas. Particle-in-cell (PIC) simulations of such sources were recently published in Refs.~\onlinecite{Hmin1, Hmin2}. Thus, the evolution of self-sustained low-pressure discharges disturbed by a high-voltage nanosecond pulse needs to be investigated at the very basic level.

In this work we performed a comprehensive study of single nanosecond pulses applied to a steady-sate low-pressure capacitively coupled RF plasma. We approached the problem experimentally, by combining time-resolved imaging of the discharge and microwave interferometry measurements. Furthermore, we supplemented our measurements with PIC simulations of our plasma. By comparing the multi-timescale evolution of the plasma in the simulations and experiments, we investigated the physical mechanisms underlying different discharge regimes.

\section{Experimental setup}
\begin{figure}[t!]
\centering
  \includegraphics [width=3.1in]{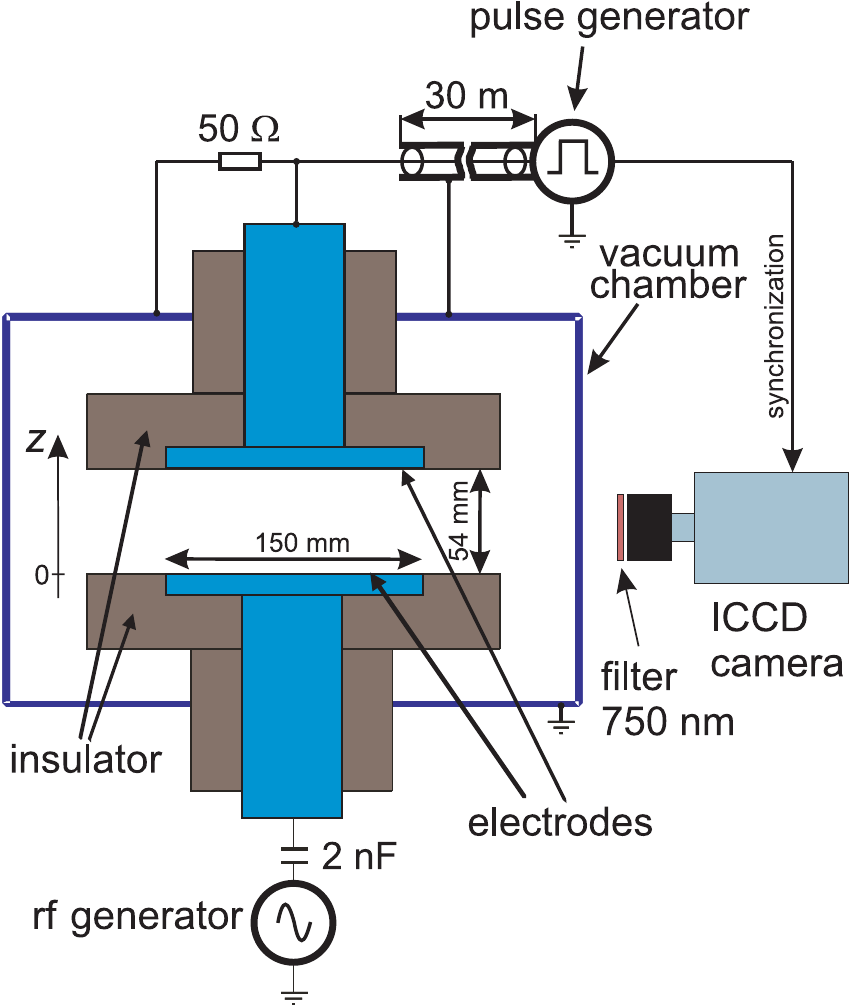}
  \caption{Experimental setup for investigations of the nanopulse discharge. High-voltage nanosecond pulses are applied to the discharge gap, in which a steady-state capacitively-coupled plasma is sustained. The $50~\Omega$ load reduces the reflection of the pulse. The ICCD camera is synchronized with the pulse generator. A $30$~m long $50~\Omega$ coaxial line delays the pulse, allowing us to observe the initial stage of the discharge development.}
  \label{Fi:STP}
\end{figure}
The experiments were performed in a parallel-plate reactor with the aluminium electrodes of $150$~mm diameter, separated by $54$~mm (Fig.~\ref{Fi:STP}). The reactor was filled with argon at a pressure $p$ of $0.1-10$~Pa. The lower electrode was connected to the RF generator via a blocking capacitor. The upper electrode was connected to the pulse generator in parallel with the $50~\Omega$ load. The RF generator continuously supplied  sinusoidal voltage at the frequency $\omega/2\pi=13.56$~MHz and peak-to-peak voltage $U_{pp}=40-100$~V to the lower electrode, producing the steady-state capacitively-coupled plasma between the electrodes. A reaction of this plasma to a high-voltage nanosecond pulse is the subject of our investigation.

Pulses from the pulse generator (FID Technologies, FPG 20-M), with the fixed duration of $20$~ns, risetime of about $2$~ns, and variable amplitude $U_A=3-17$~kV were applied to the upper electrode at the repetition frequency of $20$~Hz. The electrode design was aiming to minimize the stray capacitance, which was as low as $\approx 20$~pF and the corresponding electrode charging time $\approx1$~ns. This ensured that the electrode followed the waveform of the pulse.

The discharge gap was imaged with an Andor DH-740 ICCD camera. Opening of the image intensifier gate of the camera was synchronized with the high-voltage pulse and could be precisely positioned in time with respect to it. This allowed us to record the evolution of our discharge in a sequence of video frames using repetitive pulses. For each position of the camera gate the image was integrated on the camera CCD over $1$~s. The gate width, being the effective exposure time of each image, and step, being the effective interframe interval, were set according to the particular type of measurement. A $30$~m long $50~\Omega$ coaxial cable, through which the pulse was supplied to the chamber, served as a delay line, enabling to observe the plasma before the pulse arrival to the electrode. An interference filter with the central wavelength of $750$~nm and $10$~nm width, selecting two atomic transitions of argon, $2p_1\to 1s_2$ and $2p_5\to 1s_4$ with the lifetimes of $22.5$ and $24.9$~ns \cite{ArOpt}, respectively, was placed in front of the camera lens.

To measure the evolution of the plasma density at a late stage of the pulsed discharge we used a microwave interferometer (Miwitron, MWI 2650) with the frequency of $26.5$~GHz \cite{MWI}. Emitter was sending a probing electromagnetic wave to the plasma through a lateral glass window. The receiver was aligned with respect to the emitter in front of the opposite glass window, so that the horizontal line of sight was formed. The time-resolved phase shift $\phi$ of the probing electromagnetic wave, proportional to the line-of-sight averaged electron density $n_e$, was monitored by the oscilloscope. The time resolution of this measurement was $10$~$\mu$s. Simultaneously, we measured the intensity of integral plasma emission $I_{int}$, collecting the light from the plasma by a small collimating lens and guiding it via a $600$~$\mu$m diameter optic fiber cable to a photomultiplyer module (Hamamatsu, H7827-012) with the $200$~kHz bandwidth. The resulting curves, averaged over $32$ pulses, were recorded both for the interferometric phase shift and integral emission intensity.

\section{PIC simulations}\label{Sc:PIC}
We employed a 1D3V PIC code with Monte-Carlo collisions (MCC) \cite{Birdsall, Verboncoeur} to simulate a discharge with two parallel-plate electrodes separated by a gap of $L=50$~mm and filled with pure argon. The MCC part of the code was based on a standard approach for argon \cite{MCCArgon}.

An important modification of the standard model, caused by the need to monitor the transient processes on ns timescale, was treatment of the argon excited states which have the lifetime of the order of the pulse duration. For instance, the $1s_4$ state, which was considered in the present simulation, has the lifetime of $8.6$~ns \cite{ArRes} and the energy of transition to ground state of $11.6$~eV. The resulting vacuum ultraviolet (VUV) photons are able to produce photoemission from the electrodes with the yield $\gamma\sim 0.1$ \cite{Raizer}. Therefore, in our simulations we counted the number of $1s_4$ excited states (created by electron impact excitation of ground state atoms and decayed according to their natural lifetime). Since the plasma between the electrodes was considered to be optically thin for these VUV photons, each act of decay led with the probability $\gamma$ to immediate creation of a photoelectron at one of the electrodes. Given the short lifetime of the excited states, thus allowed us to discard their spatial distribution and only account for their total number.

Similar to the experiment, in our simulations we first generated a steady-state discharge. We set appropriate boundary conditions on the electrodes, i.e. a sinusoidal voltage of $13.56$~MHz frequency was applied to one electrode and the other electrode was grounded. After the RF discharge reached equilibrium, a high voltage was applied to the previously grounded electrode during the period of $\tau=20$~ns and then grounding was restored again. Subsequent relaxation of the discharge was monitored.

In the experiment, ICCD camera registered the light emission intensity $I_{exp}$, whereas the simulation dealt with the plasma kinetics and therefore allowed us to access the excitation rate. Evolution of the emission intensity is determined by the convolution of the excitation rate $\Gamma_{exc}(t)$ and the exponential decay function $\exp{(-t/T)}$, where $T$ is the lifetime of the upper level of the transition. In order to compare the simulation and experimental results, we therefore recalculated simulated excitation rate into the emission intensity $I_{sim}$, using $T=24.9$~ns for the lifetime of the $2p_5\to 1s_4$ transition.

\section{Experimental results}
\begin{figure*}[t!]
\centering
  \includegraphics [width=6in]{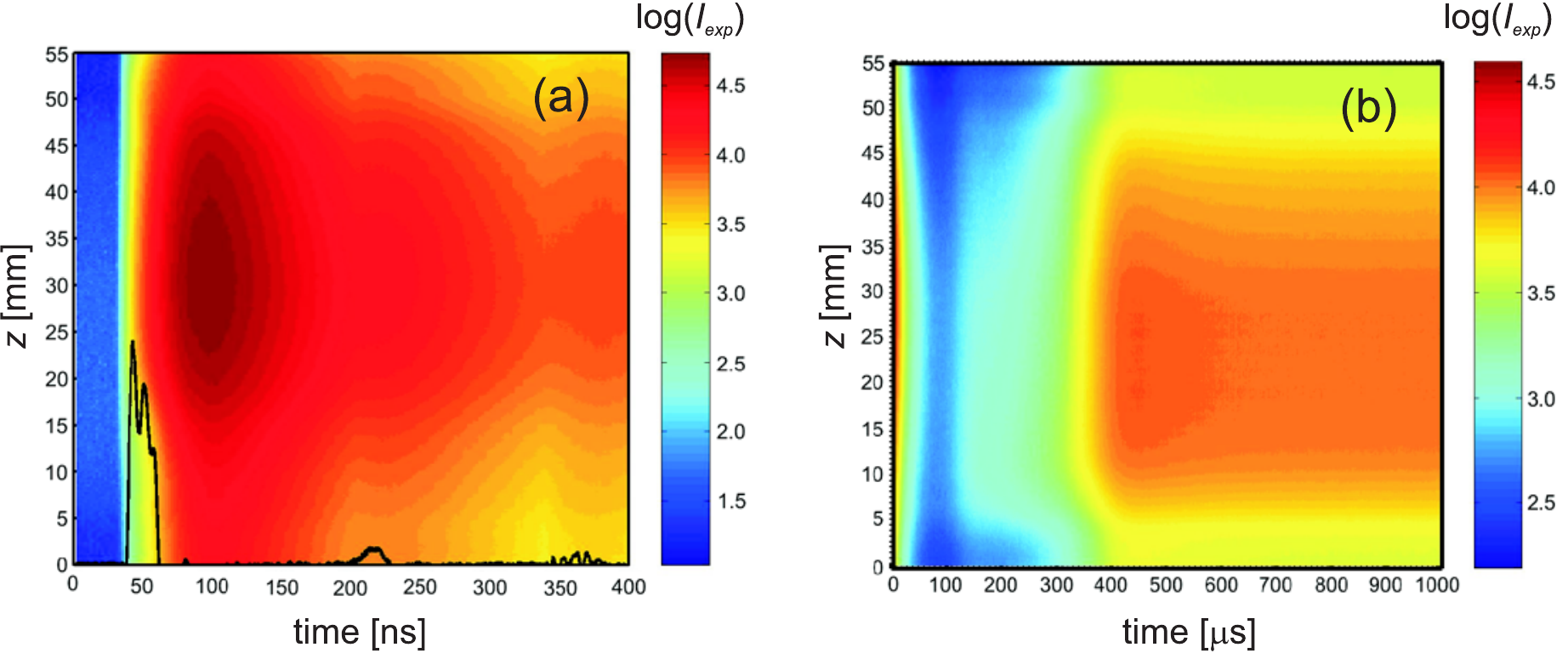}
  \caption{(a) Space-time diagram of the flash (ICCD gate width $10$~ns, gate step $2$~ns). The black line shows the evolution of the voltage on the upper electrode. An afterpulse (at $\simeq200$~ns) and a re-reflected pulse (at $\simeq350$~ns) give rise to additional emission peaks. (b) Space-time diagram of the dark phase (ICCD gate width $2~\mu$s, gate step $2$~$\mu$s). The shown results are for $p=3$~Pa, $U_{pp}=100$~V and $U_A=8$~kV. For each value of $z$ the intensity is averaged over approximately $4$~cm horizontally. Note that intensities in the two panels cannot be directly compared due to different gate widths.}
  \label{Fi:XTs}
\end{figure*}

The plasma relaxation after the high-voltage pulse turned out to be quite a complicated multi-timescale process with two characteristic regimes: A bright flash at the initial stage of the discharge with the characteristic width of the order of $100$~ns (when emission intensity increases $2-3$~orders of magnitude above the steady-state level), and the so-called dark phase lasting from several hundreds of $\mu s$ to several ms (when the emission intensity drops $1-2$~orders of magnitude below the steady-state level). The latter regime appears to be similar to that reported in Refs. [\onlinecite{Amirov1}-\onlinecite{DP2}]. Typical space-time diagrams for both regimes are presented in Fig.~\ref{Fi:XTs}a and~\ref{Fi:XTs}b.

We note that the flash does not occur during the pulse, like in pulsed discharges at atmospheric pressure \cite{KorMes}. Significant growth of emission intensity starts when the high voltage is removed from the electrode.

\begin{figure}[t!]
\centering
  \includegraphics [width=3.1in]{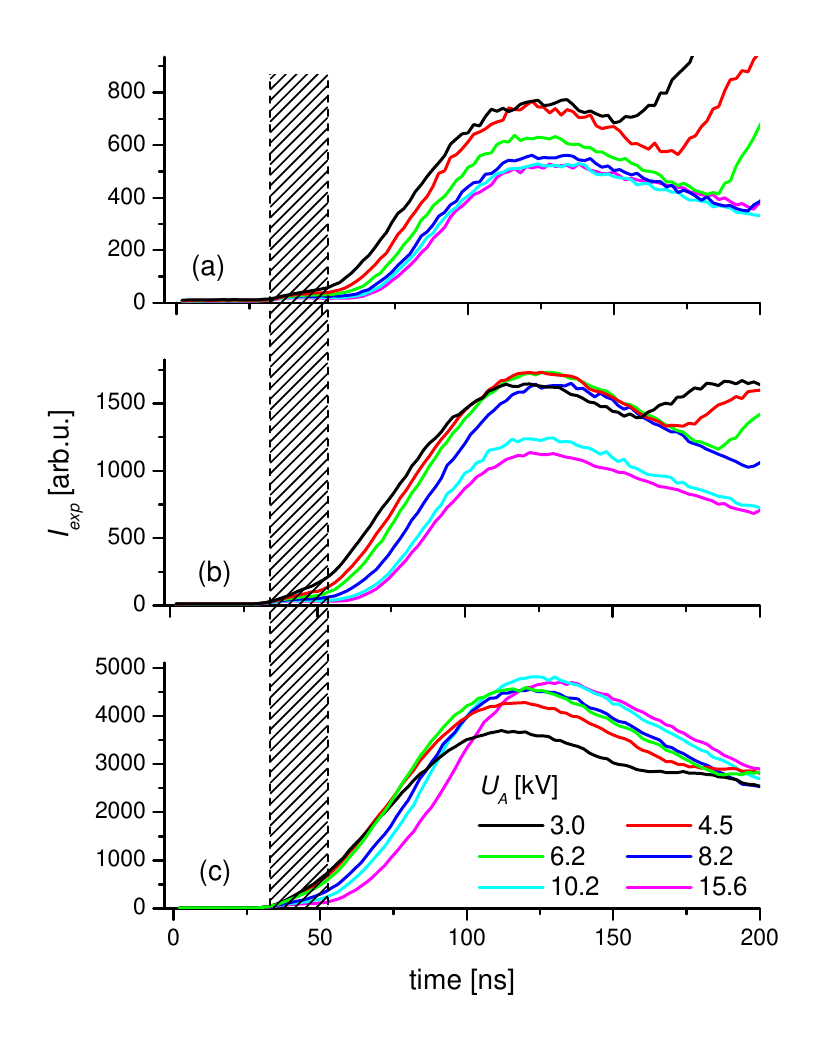}
  \caption{Temporal evolution of the light emission at the initial (flash) stage, for $p=1.5$~Pa and (a) $U_{pp}=40$~V, (b) $U_{pp}=56$~V, (c) $U_{pp}=100$~V. Growth of the emission intensity after the peak is associated with the afterpulse (Fig.~\ref{Fi:XTs}a). The dashed area indicates the high-voltage pulse.}
  \label{Fi:IUppUA}
\end{figure}

The light emission at the flash stage is characterized by a complicated dependence on $U_A$. We compared these dependencies measured for three different values of $U_{pp}$. For the smallest $U_{pp}=40$~V (Fig.~\ref{Fi:IUppUA}a) the intensity primarily decreases with $U_A$, for $U_{pp}=56$~V (Fig.~\ref{Fi:IUppUA}b) it first increases and then decreases, and for $U_{pp}=100$~V (Fig.~\ref{Fi:IUppUA}c) it primarily increases. This is another distinct feature of our discharge. Usually, in high-pressure pulsed discharges \cite{KorMes} the flash intensity monotonously increases with the pulse amplitude.

We note here that in our experiments the emission intensity exhibits a series of flashes (rather than a single initial flash), as can be seen in Fig.~\ref{Fi:XTs}a. The same is evident in Fig.~\ref{Fi:IUppUA}, where the intensity starts growing again after the first flash. These ``follow-up flashes'' occur due to the afterpulse (with amplitude $\simeq20\%$ of $U_A$) produced by the pulse generator, as well as due to the re-reflection of the main pulse. Nevertheless, in all cases the initial flash is well separated and its intensity can be easily determined.

Unfortunately, the presence of follow-up flashes did not allow us to perform careful studies of the effect of $U_A$ on the dark phase.
\begin{figure}[t!]
\centering
  \includegraphics [width=3.1in]{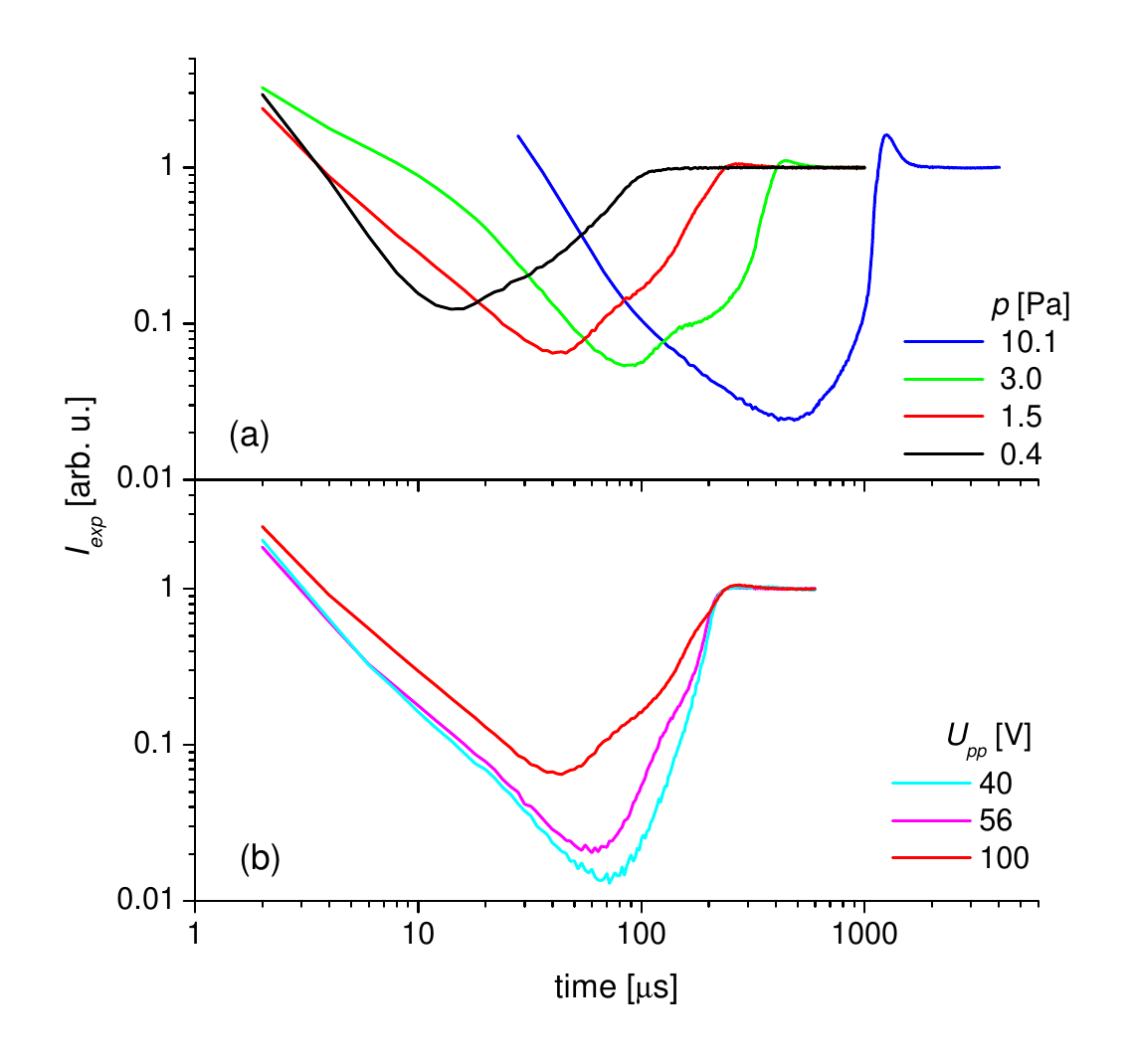}
  \caption{Temporal evolution of the light emission during the dark phase, obtained for $U_A=8$~kV. The figure shows (a) curves for different $p$ at fixed $U_{pp}=100$~V and (b) curves for different $U_{pp}$  at fixed $p=1.5$~Pa. All curves are normalized to the steady-state emission intensity.}
  \label{Fi:DPExp}
\end{figure}
In our experiments the dark phase was observed for practically all studied plasma conditions, as illustrated in Fig.~\ref{Fi:DPExp}. With the increase of pressure an overshoot of emission intensity (obvious also in a space-time diagram on Fig.~\ref{Fi:XTs}b) starts to develop at the end of the dark phase.

\begin{figure}[t!]
\centering
  \includegraphics [width=3.1in]{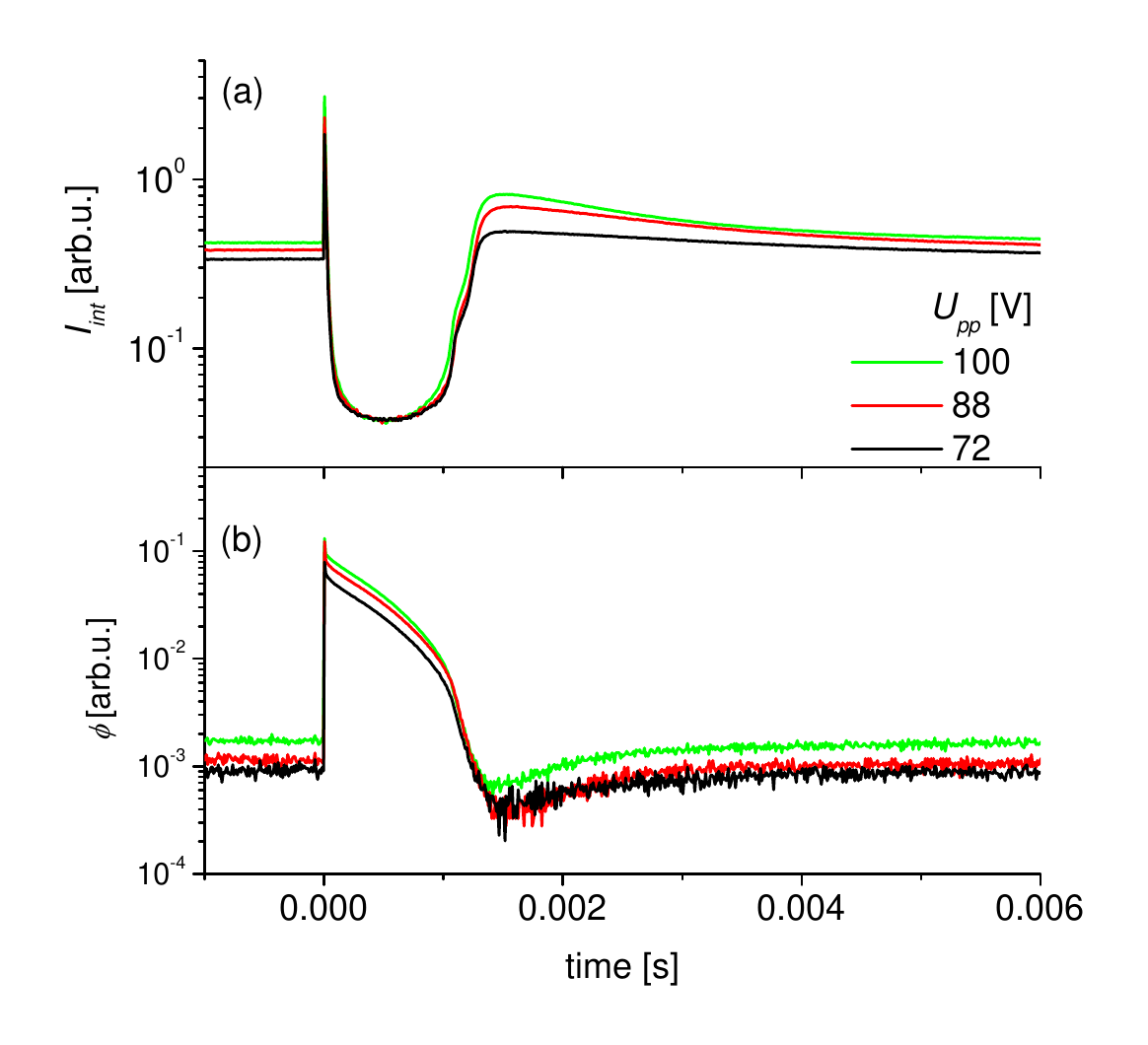}
  \caption{Temporal evolution (a) of the integral emission intensity and (b) of the phase shift in the microwave interferometry for $p=10$~Pa and $U_A=8$~kV. The dark phase is accompanied by a dramatic increase of the plasma density.}
  \label{Fi:MWI}
\end{figure}

The results of the microwave interferometry measurements for $p=10$~Pa are shown in Fig.~\ref{Fi:MWI}. They indicate that the plasma density, tremendously increased during the flash, remains very high (compared to its steady-state value) also during the dark phase. Measurements at smaller pressures exhibit similar dynamics of plasma density in the dark phase, whereas the steady-state values are too small to be measured reliably.
The importance of ``high-pressure'' measurements shown in Fig.~\ref{Fi:MWI} is that they allow us to track the variation of plasma density also in the overshoot. They clearly demonstrate that during the overshoot $n_e$ drops below the steady-state value.

\section{Discussion}
In order to identify the physical mechanisms underlying the observed behavior of the plasma, we compare our experimental results with the results of PIC simulations. We do it separately for the flash and the dark phase regimes.
\subsection{Flash}\label{Sc:IEP}
Although the profile of the high-voltage pulse is not free from some spurious features seen in Fig.~\ref{Fi:XTs}a, the resulting flash, as mentioned above, is always easy to identify. Therefore, below we discuss the flash assuming that it was created by a single high-voltage pulse of $20$~ns duration and given amplitude. For our PIC simulations (Sec.~\ref{Sc:PIC}) we used pulses of this idealized shape.
\subsubsection{Mechanism of flash generation} \label{Sc:QA}
\textit{Qualitative analysis.} Before starting a detailed comparison between our experiments and simulations, we shall demonstrate that even a simple approach based on elementary estimates and scalings can explain the main characteristics of the flash regime.

Let us consider a quasineutral plasma slab with the density $n$ between two infinite plain electrodes, separated by a gap $L$. At a certain moment the voltage $U_A$ is applied to one of the electrodes for the period $\tau$. In this consideration we completely neglect the RF electric field (which sustained the steady-state discharge), since it is supposed to be much smaller than the pulse field $E_A=U_A/L$. After the pulse field is applied, electrons in the plasma start moving. For $p\sim1$~Pa, $L=50$~mm and $U_A\sim1$~kV the electron-neutral collisions can be neglected and the electron motion can be considered as ballistic. Time required for an electron to cross the gap is then
\begin{equation}
\tau_e=L\sqrt{\frac{2m_e}{eU_A}},
\end{equation}
where $m_e$ and $e$ are the electron mass and charge, respectively. For the range of $U_A$ employed in our experiments $\tau_e$ lies between $1$ and $3$~ns and therefore is much smaller than the pulse time $\tau$. Hence, electrons are able to leave the gap during the pulse.

As electrons are leaving the plasma, the bulk positive charge due to immobile ions (their characteristic time of flight $\tau_i=\tau_e\sqrt{m_i/m_e}$ is in the sub-$\mu$s range) starts building up. Therefore, after removal of the pulse, the \textit{``residual''} electric field $E_{res}$ is generated: It is determined by $\partial E_{res}/\partial z\sim e(n-n_{res})/\epsilon_0$, where $z$ is the discharge axis and $n_{res}$ is the \textit{residual} electron density (remaining in the discharge gap after the removal of the pulse field). Hence we get the following estimate for the ``residual'' field:
\begin{equation}
E_{res}\sim\frac{e(n-n_{res})L}{\epsilon_0}.
\label{Eq:ResE}
\end{equation}
We conclude that the ``residual'' electric field can vary in the range $0<E_{res}\lesssim E_i$, where
\begin{equation}
E_i=\left. E_{res}\right|_{n_{res}=0}\sim\frac{enL}{\epsilon_0}
\end{equation}
is the electric field of immobile bulk ions. For a typical plasma density $n=2\times10^{14}$~m$^{-3}$ we get $E_i\sim2\times10^5$~V/m.

Thus, after removal of the pulse field, residual electrons get accelerated by the ``residual'' field and start  ionizing the neutral gas. Significant electric field will therefore be present until the excess ions are diluted by the newly generated electrons and ions. Since the ``residual'' field is due to positive bulk charge, electrons are trapped inside the gap, which provides ideal conditions for the ionization boost.

Now let us qualitatively consider the dependence of the flash intensity on the pulse field $E_A$. When the pulse field is sufficiently small, $E_A\ll E_i$, it is effectively screened by the plasma, so that $E_{res}\sim E_A$. This naturally causes the flash intensity to grow with $E_A$. On the other hand, for $E_A\gg E_i$ the plasma cannot provide the pulse screening, since the ``residual'' field is limited by $E_i$. In this case, there are (practically) no electrons left in the discharge gap after the pulse~\textendash\, electrons have to be first generated before they start effectively ionizing and exciting neutral gas. This implies that the flash intensity reaches maximum at some $E_A\lesssim E_i$, i.e, it is a non-monotonic function of the ratio $E_A/E_i$.

This non-monotonic dependence of the flash intensity on $E_A$ was observed in the experiment, as illustrated in Fig.~\ref{Fi:IUppUA}. The used values of $U_A$ provided the variation of $E_A$ in the range of $(0.6-3.4)\times10^5$~V/m, which was sufficient to observe evidence of both increase and decrease of the flash intensity at fixed plasma conditions (Fig.~\ref{Fi:IUppUA}b). Moreover, by varying $U_{pp}$ and, in this way extending the range of $E_A/E_i$, (since $E_i$ grows with plasma density which, in its turn, grows with $U_{pp}$) we were able to achieve the regimes of the major increase (Fig.~\ref{Fi:IUppUA}c) and decrease (Fig.~\ref{Fi:IUppUA}a) of the flash intensity with $U_A$, occurring at higher and lower values of the plasma density, respectively.

The flash is therefore caused by the electric field of bulk ions, which remain uncompensated for a short time after the high-voltage pulse. Possibility of such mechanism of discharge ignition was discussed in Ref.~\onlinecite{SchneiderTVT}. For instance, similar transient decompensation can ignite discharges in solid dielectric materials irradiated by pulsed electron beams of MeV energy \cite{MeVElectr}. Also, the so-called transient luminous events (TLEs) in the upper Earth atmosphere occur as a result of such a decompensation caused by lightning \cite{TLE}.

\begin{figure}[t!]
\centering
  \includegraphics [width=3.1in]{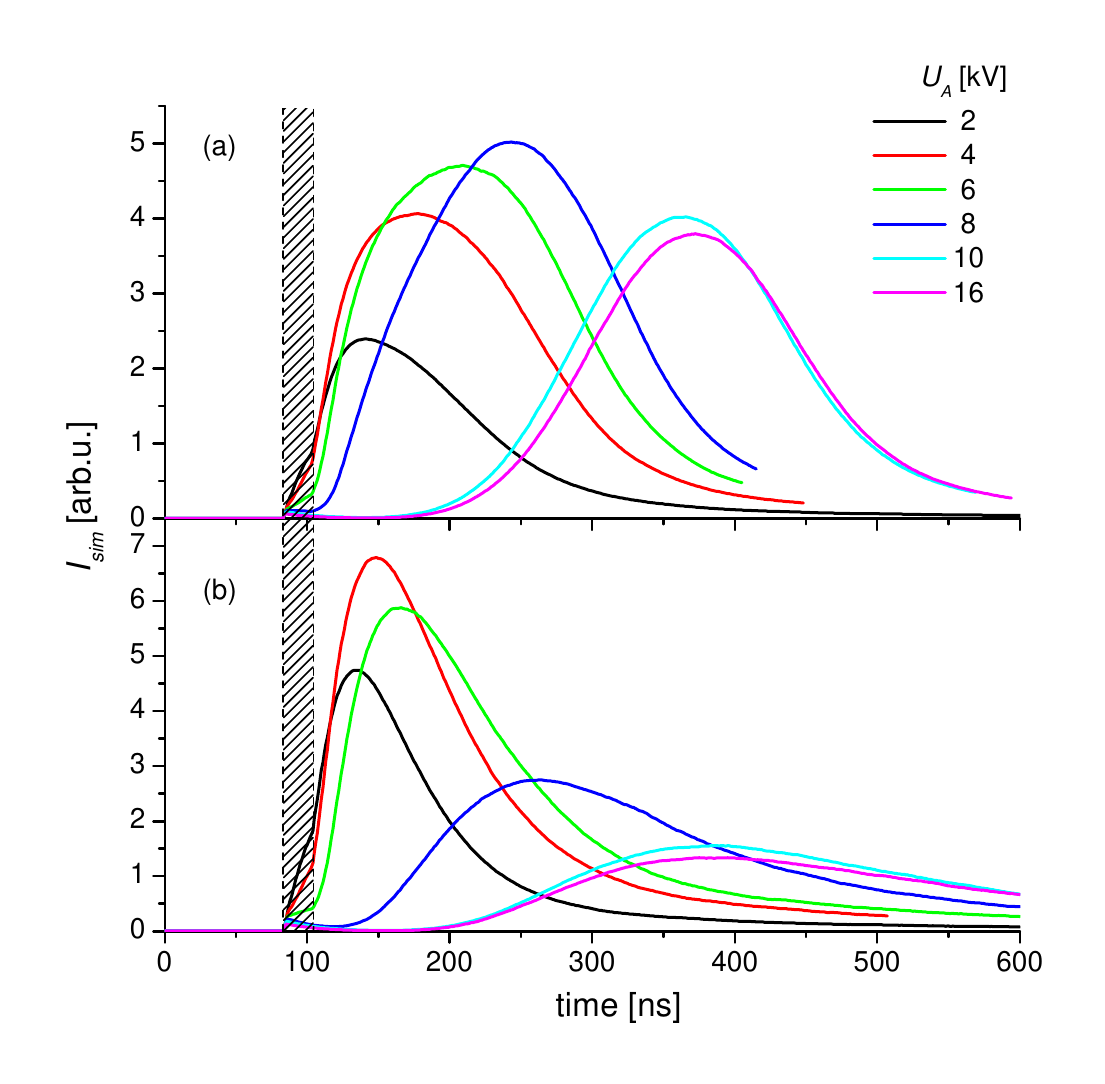}
  \caption{Simulated behavior of the emission peak during the flash stage, demonstrating the effect of the secondary electrons for (a) $\gamma=0.1$ and (b) $\gamma=0$. The results are for $p=3$~Pa, $U_{pp}=100$~V and different values of $U_A$. The dashed area indicates the high-voltage pulse.}
  \label{Fi:ISim}
\end{figure}
\textit{Comparison with PIC simulations.} Our simulations confirm main qualitative findings. Figure~\ref{Fi:ISim}a presents the dependence of the simulated flash intensity on $U_A$. Its non-monotonic character becomes evident for $U_A$ between $4$ and $16$~kV. In the experiment, as we already mentioned, this range of $U_A$ was insufficient to clearly demonstrate the non-monotonic dependence. This suggests that steady-state plasma density in the experiment was higher than in simulation.
\begin{figure}[t!]
\centering
  \includegraphics [width=3.1in]{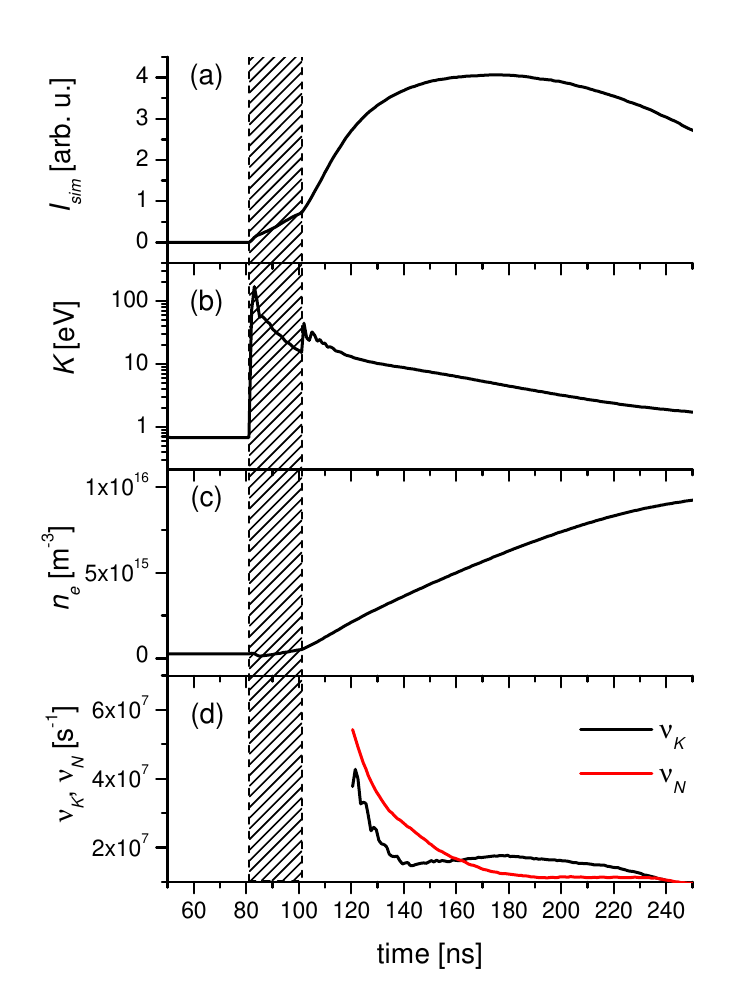}
  \caption{Simulated evolution of plasma parameters (for $p=3$~Pa, $U_{pp}=100$~V, $U_A=4$~kV), showing the formation of the emission peak at the flash stage: (a) emission intensity $I_{sim}$, (b) mean electron energy $K$, (c) electron density $n_e$, (d) relaxation rates $\nu_K=\left|K^{-1}\partial K/\partial t\right|$ and $\nu_N=\left|N^{-1}\partial N/\partial t\right|$, where $N$ is determined by Eq.~(\ref{Eq:N}). The dashed area indicates the high-voltage pulse.}
  \label{Fi:PF}
\end{figure}

Figure~\ref{Fi:PF} illustrates formation of the first emission peak. Here we discuss a particular example of a relatively small $E_A$ ($U_A=4$~kV), when significant fraction of electrons is left in the discharge gap after the pulse. However, the same physics is also valid in a case of strong $E_A$. The pulse field and, later, the ``residual'' field (see two respective peaks in Fig.~\ref{Fi:PF}b) accelerate electrons to energies $K\sim100$~eV. After that electrons start loosing energy. At the same time, the number of electrons grows (Fig.~\ref{Fi:PF}c). Remarkably, the emission peak occurs very close to the local maximum of the electron energy loss rate, $\nu_K=\left|K^{-1}\partial K/\partial t\right|$ (compare Fig.~\ref{Fi:PF}a and Fig.~\ref{Fi:PF}d), suggesting that the main mechanism of electron energy loss are inelastic collisions. Therefore, one can argue that the emission peak occurs when the energy spectrum of electrons (which are instantaneously cooled down) becomes ``optimized'' for the impact excitation (whose cross section is a non-monotonic function of electron energy). Delay of the emission peak seen in simulations at higher $U_A$ (Fig.~\ref{Fi:ISim}a) is then explained by much higher initial electron energy (due to much larger $E_A$ and $E_{res}$). We notice, however, that the position of the emission peak has a significantly different tendency with $U_A$ in the experiment, where it is practically independent of $U_A$.

Figure~\ref{Fi:PF}d shows the behavior of two logarithmic derivatives, $\nu_K$ and $\nu_N=\left|N^{-1}\partial N/\partial t\right|$, where
\begin{equation}
N=\frac{n_i-n_e}{n_e}
\label{Eq:N}
\end{equation}
is the relative density disparity, with $n_e$ and $n_i$ being the momentary electron and ion densities, respectively. The parameter $\nu_N$ reflects the decay rate of the ``residual'' field. We notice that $\nu_N>\nu_K$ during a certain period of time before the emission peak (from $110$~ns to $160$~ns). This suggests that even in a 1D case some diffusion cooling takes place: as the newly generated electrons and ions appear and dilute the excessive ion charge, the most energetic electrons become no longer trapped and leave the plasma. In 1D simulations the electrons can only be lost at the electrode surfaces, whereas in a real 3D discharge a significantly larger fraction of electrons can leave the discharge volume in the lateral direction and also be lost on the surfaces of dielectric insulators. Therefore, in a 3D discharge a higher diffusion cooling rate is expected. These additional losses can, in principle, reduce sensitivity of the position of the emission peak to the initial electron energy spectrum.

\subsubsection{Role of secondary electrons}
In Sec.~\ref{Sc:PIC} we pointed out that surface electron production may play an important role at the initial stage of the discharge development. For large $E_A$ and small $n_{res}$, the electron impact ionization is suppressed right after the pulse, and therefore electron emission from surfaces may become essential for the further evolution. In our case, the secondary electron emission can be produced by ions, neutral metastable particles, electrons, and photons. Since heavy particles are too slow, they can be \textit{a priori} excluded from the consideration on ns timescales. Electron-electron emission can also be ruled out because the typical yield for metallic electrodes never exceeds unity. At the same time, high-energy photons, being fast and insensitive to the electric field, certainly can affect the processes evolving on ns timescales.

There are two sources of high-energy photons in our discharge: (i) bremsstrahlung generated during the pulse by energetic (keV) electrons hitting the electrode surface and (ii) VUV light emitted by resonant transitions in Ar atoms. Small electron-photon conversion efficiency ($\sim10^{-5}$ \cite{XRay}) suggests that bremsstrahlung photons cannot play any important role. As regards the effect of the resonance states of argon, it was taken into account in our simulations.

Figure~\ref{Fi:ISim} shows the evolution of the plasma emission at the initial flash phase calculated for two different values of the photoemission yield. One can see that for $\gamma=0$ the peak emission intensity retains the same non-monotonic dependence on $U_A$ as in the case $\gamma=0.1$. However, after the maximum is reached, the peak intensity drops with $U_A$ much faster in the case $\gamma=0$. This effect can be easily understood: During the pulse a certain number of the resonant Ar states is generated in the discharge. Since their decay time ($8.6$~ns) is comparable to the pulse duration, they primarily remain in the discharge gap after the pulse and supply the plasma with additional electrons. This leads to effective increase of the residual electron density and, hence, increase of the flash intensity. The effect of Ar resonant states slowly decreases with $U_A$, since the excitation cross-section decreases with the electron energy (in the keV range).

\subsection{Dark phase}\label{Sc:DP}
\begin{figure}[t!]
\centering
  \includegraphics [width=3.1in]{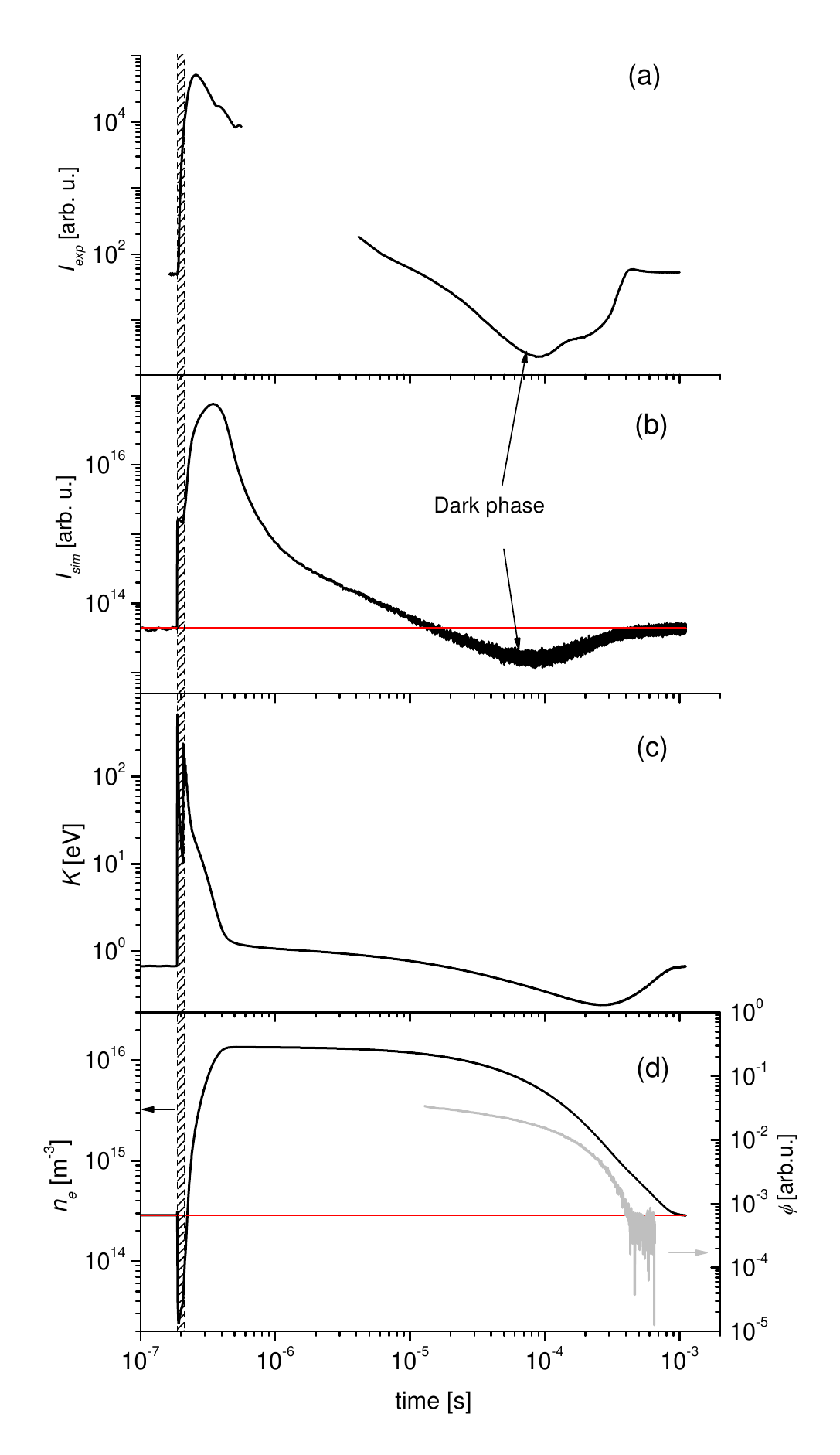}
  \caption{Evolution of (a) experimentally measured light emission (combination of central cross-sections of Figs.~\ref{Fi:XTs}a and~\ref{Fi:XTs}b) and simulated (b) emission intensity, (c) average electron energy and (d) electron density. In (d), also the experimental microwave interferometric phase shift $\phi\propto n_e$ is shown (steady state value, is close to the limit of detection and therefore is noisy). The shown results are for $U_{pp}=100$~V, $p=3$~Pa and $U_A=8$~kV. The \emph{steady-state discharge} is characterized by relatively small plasma density and low average energy of electrons. \emph{During the pulse} the electron density is drastically decreased, since electrons are swept away by the pulse field. When the high voltage is removed, the plasma density (see also Fig.~\ref{Fi:MWI}b) and average electron energy increase and acquire the values much higher than those in a steady-state plasma (see Sec.~\ref{Sc:IEP} and Fig.~\ref{Fi:PF}) This occurs within a fraction of $\mu$s due to the presence of a strong ``residual'' field. The subsequent relaxation to the steady-state discharge is accompanied by the \emph{dark phase} lasting a fraction of ms (see Sec.~\ref{Sc:DP}). The dashed area indicates the high-voltage pulse.}
  \label{Fi:SLT}
\end{figure}
Figures~\ref{Fi:SLT}a and~\ref{Fi:SLT}b present a typical multi-timescale evolution of the discharge emission in our experiments and simulations, respectively. A remarkable qualitative agreement is observed: the dark phase~\textendash~its position and duration~\textendash~are well reproduced in the simulations. High values of the plasma density during the dark phase (Fig.~\ref{Fi:MWI}) are obtained in the simulated discharge as well (compare evolution of $\phi$ and $n_e$ in Fig.~\ref{Fi:SLT}d).

We see that a tremendous increase of $n_e$ is accompanied by a steep reduction in the average electron energy (Fig.~\ref{Fi:SLT}c). By comparing Figs.~\ref{Fi:SLT}b and~\ref{Fi:SLT}c we notice that the minimum of emission intensity occurs about $200$~$\mu$s before the minimum of $K$, whereas the electron density is monotonically decreasing during the dark phase (Fig.~\ref{Fi:SLT}d). Obviously, the excitation processes (reflected by the emission intensity) are governed by high-energy tail of the electron distribution function whose kinetics can differ significantly from that of the mean energy $K$.

We note that in simulations the RF peak-to-peak voltage was kept constant, whereas in experiments, where the discharge is a part of a real electrical circuit, steadiness of the peak-to-peak voltage cannot be guaranteed: a dramatic increase of the electron density inside the discharge gap causes significant drop of the discharge active resistance, leading to voltage redistribution in the circuit. Such a mechanism is responsible for the formation of a dark phase during the ignition of dc glow discharges [\onlinecite{Amirov1}-\onlinecite{DP2}]. Our experiment is also not completely free from this effect. We found that after the pulse the RF peak-to-peak voltage on the electrode typically experiences a $15\%$ drop. This drop, however, is hardly enough to reduce the emission intensity even by a factor of two, whereas Figs.~\ref{Fi:SLT}a and~\ref{Fi:SLT}b show the reduction by an order of magnitude and more.

Our simulations clearly suggest a different mechanism which leads to the electron cooling in the high-density capacitively-coupled RF plasma. To understand this mechanism, let us consider how the dielectric permittivity of a collisionless plasma, $\epsilon=1-(\omega_p/\omega)^2$ (where $\omega_p\sim n_e^{1/2}$ is the electron plasma frequency) varies after the pulse. Already in steady-state conditions $\omega_p/\omega\approx 10$, so that $\epsilon$ is strongly negative and the discharge is stabilized by a weak screening of the RF  field. As the electron density grows after the pulse, the screening becomes even stronger. In this regime the penetration depth of the RF field into the plasma is $\delta\simeq c/\omega_p$ \cite{RFRaizer}. For the peak electron density $\sim10^{16}$~m$^{-3}$ (see Fig.~\ref{Fi:SLT}d) $\omega_p/\omega\approx 80$ and $\delta\approx5.3$~cm, which is very close to the interelectrode separation $L$. This indicates that the RF field at this stage is effectively screened by the high density plasma, and electrons can cool down. As the significant fraction of excess charges leave the plasma due to the ambipolar diffusion, RF field starts penetrating deeper into it and heats up electrons again. In the electrotechnical sense, the discharge gap becomes strongly inductive during the dark phase and, therefore, efficiently reflects the RF power.

In our experiment the plasma relaxation was sometimes accompanied by two features which we were not able to reproduce in simulations: a ``knee'' following the bottom of the dark phase and an ``overshoot'' at its end (Fig.~\ref{Fi:SLT}a). A hint about the origin of the knee can be found in Ref.~\onlinecite{Aglow}, where similar features occurring at similar timescales were observed during the afterglow of the inductively-coupled low-pressure RF discharge in argon. The authors explained the observed features by the decay of excited atoms created as a result of three-body recombination. In our case, the electron impact excitation rate drops significantly during the dark phase and, therefore, contribution of the recombination (which is not taken into account in our simulations) to the population of 2p levels might indeed be significant.

Microwave interferometry measurements performed at higher pressures (Fig.~\ref{Fi:MWI}) deliver further information on the overshoot and provide additional support for the suggested mechanism of the dark phase formation. The plasma density during the overshoot is reduced below the steady-state level, which weakens the screening of RF field and, therefore, leads to more efficient heating of electrons. This, however, should not be understood as a self-consistent explanation of the observed overshoot since the reason for the reduction of electron density remains unknown. Refs.~\cite{DP1, DP2} attribute similar features to the kinetics of metastable atoms, which is not considered in our simulations.

\section{Conclusion}
Our experiments and simulations showed that a high-voltage nanosecond pulse applied to a steady-state capacitively-coupled low-pressure weakly ionized plasma produces a profound long-lasting disturbance. The resulting effects are governed by a variety of different mechanisms operating in plasma at essentially different time scales. One can identify two principal regimes: the flash, lasting about a hundred ns (up to several hundreds ns in the simulations) after the pulse, and the dark phase, with the duration from a few hundreds $\mu$s to a few ms.

Since the pulse field is comparable to the electric field of bare ions in the steady-state plasma, a significant fraction of electrons is swept away from the discharge gap during the pulse. Directly demonstrated in simulations, this effect also found an indirect confirmation in the experimental dependencies of the flash intensity on the pulse amplitude: the intensity increased with the pulse amplitude at higher densities of the steady-state plasma, whereas at lower densities it starts decreasing.

After the pulse field is removed, the residual electrons start accelerating in a strong field of immobile (at this stage) ions, which generates the flash. We showed that the flash intensity is maximal when the pulse field is somewhat smaller than the field of (immobile) bare ions. Secondary electron emission due to VUV radiation from the excited argon states turned out to be important at this stage.

During the dark phase following the bright flash, the emission intensity drops below the steady-state value. Both the simulations and time-resolved microwave interferometry measurements showed that the electron density during this phase is much higher than that in the steady-state plasma. In such a dense plasma the penetration depth of the RF field decreases and becomes comparable to the interelectrode gap. This screening effect leads to effective cooling of electrons and the subsequent decrease of the emission intensity.

Thus, 1D3V PIC simulations provide good qualitative explanation of the major features observed in our experiments. Investigation of unresolved issues, such as the effect of the pulse amplitude on the flash delay or the origin of the knee and overshoot seen in emission intensity during the dark phase, require additional careful experiments and numerical simulations.

\section{Acknowledgements}
This work was partially carried out within the framework of the European Fusion Development Agreement and the French Research Federation for Fusion Studies.

\end{document}